\documentclass[conference]{IEEEtran}
\IEEEoverridecommandlockouts
\PassOptionsToPackage{table,dvipsnames}{xcolor}

\usepackage[english]{babel}
\usepackage{ragged2e}
\usepackage{caption}
\usepackage[table]{xcolor}
\usepackage{siunitx}
\usepackage{enumitem}
\usepackage[final,expansion=false]{microtype}
\usepackage[hidelinks]{hyperref}
 
\definecolor{darkblue}{HTML}{1F3864}

\usepackage[T1]{fontenc}
\usepackage[utf8]{inputenc}
\usepackage{cite}
\usepackage{amsmath,amssymb,amsfonts}
\usepackage{graphicx}
\usepackage{textcomp}
\usepackage{url}
\usepackage{booktabs}
\usepackage{tabularx}
\usepackage{multirow}
\usepackage{array}
\usepackage{colortbl}
\usepackage{tikz}
\usetikzlibrary{arrows.meta, positioning, shapes.geometric, fit, backgrounds}
\usepackage{makecell}
\usepackage{orcidlink}
\usepackage{pgfplots}
\pgfplotsset{compat=1.17}

\newcolumntype{C}[1]{>{\centering\arraybackslash}p{#1}}
\newcolumntype{L}[1]{>{\raggedright\arraybackslash}p{#1}}

\definecolor{headerBlue}{RGB}{31,73,125}
\definecolor{rowGray}{RGB}{242,242,242}

\newcommand{\hdr}[1]{{\bfseries #1}}

\colorlet{headblue}{headerBlue}
\colorlet{rowgray}{rowGray}

\begin{document}

\title{Perceived Effectiveness of Social Debt Mitigation Measures: A Preliminary Vignette-Based Survey}

\author{
\IEEEauthorblockN{Philipp M. Zähl~\orcidlink{0000-0003-3302-4415}, Tom Siegel, Robin Kha, Chaitanya Chikkala, Zoé Lermusiaux~\orcidlink{0009-0000-0736-3566}, and Anika Hennig~\orcidlink{0009-0001-9418-8512}}
\IEEEauthorblockA{iDA Institute for Digitalization\\
FH Aachen University of Applied Sciences\\
Aachen, Germany\\
\{zaehl, lermusiaux, hennig\}@fh-aachen.de}

}

\maketitle
\thispagestyle{plain}
\pagestyle{plain}

\begin{abstract}
Software engineering is a socio-technical discipline in which unresolved social and organizational problems accumulate as \emph{social debt}, gradually eroding collaboration and project health. The literature offers a solid vocabulary for these problems but says little about which countermeasures practitioners consider effective. We address this gap with a vignette-based survey: six social debt problem dimensions (Organizational Silo, Newcomer Isolation, Communication Breakdown, Knowledge Bottleneck, Cultural and Geographical Distance, Psychological Safety Deficit) were presented as realistic team scenarios, and 23 IT professionals and advanced students rated 41 literature-derived mitigation measures. Rankings were analyzed with paired $t$-tests and a hierarchical Bayesian model that accounts for the within-subject design and yields, for each measure, the probability of being the best option in its scenario. Only for Communication Breakdown is the answer unambiguous: consolidating communication into a single primary channel is almost certainly preferred ($P(\text{best})=0.91$). For the other five dimensions, the model identifies favorites with moderate evidence ($P(\text{best})$ between 0.38 and 0.58), such as structured knowledge transfer and blameless retrospectives. Across scenarios, structural measures that change how information, decisions, and knowledge are organized were rated higher than role-based or attitudinal measures (by 0.3 to 0.4 scale points), although not in every scenario. Given the small convenience sample, the results are exploratory, but they offer software engineering managers an evidence-graded starting point for selecting mitigation measures.
\end{abstract}



\section{Introduction}

Software engineering is a socio-technical discipline: delivering software depends not only on technical competence but also on how people communicate, coordinate, and collaborate \cite{tamburri2013,tamburri2019}. Communication breakdowns, fragmented collaboration structures, and interpersonal friction rarely cause immediate failure. Instead, they accumulate quietly and erode software quality and organizational effectiveness over time \cite{caballero2023,zahl2026}. Tamburri et al.\ capture this accumulation as \emph{social debt}, ``an additional project cost'' caused by suboptimal social and organizational interactions, analogous to technical debt but rooted in human rather than technical compromises \cite{tamburri2015}. Its reported consequences include reduced productivity, knowledge bottlenecks, lower psychological safety, and long-term collaboration difficulties \cite{ahmad2022,saeeda2024b,dreesen2021}.

The problem has grown with the complexity of modern development settings. Globally distributed teams, agile delivery structures, and tightly coupled socio-technical systems demand continuous coordination \cite{lambiase2022,dreesen2021}. At the same time, the dysfunctions that produce social debt, such as siloed sub-teams, dependence on single individuals, weak onboarding, or fear of disagreement, tend to remain invisible in everyday work because they are informal and gradual \cite{caballero2023,ahmad2023}. Social debt is therefore often recognized only in retrospect \cite{zahl2026}.

The literature now offers a mature vocabulary for describing these problems, but comparatively little guidance on how to address them. Organizations adopt cross-team forums, psychological-safety initiatives, or knowledge-sharing programs, yet, there is little empirical evidence about which measures practitioners consider most effective for which problems. Generic interventions risk missing the underlying cause and wasting organizational resources. Existing work is largely qualitative and conceptual \cite{caballero2023,zahl2026}; it identifies candidate countermeasures but does not compare them against each other.

To address this gap, we asked IT professionals and advanced IT students to rate literature-derived mitigation measures for six social debt problem dimensions, each presented as a realistic team scenario (vignette). This allows measures to be compared against their direct alternatives for the same problem rather than in the abstract. The study is guided by the following research question:

\medskip
\noindent\textit{RQ: Which social debt mitigation measures do IT professionals and advanced students perceive as most effective for specific social debt problem dimensions?}
\medskip

The study makes three contributions. First, it translates qualitatively identified social debt dimensions into structured vignettes suitable for quantitative evaluation. Second, it provides a comparative assessment of 41 mitigation measures across six dimensions and quantifies how much confidence each ranking deserves. Third, it tests whether a recurring pattern across scenarios, namely a preference for structural over attitudinal measures, holds up when examined directly and derives from this a practitioner-oriented prioritization.

Section~\ref{sec:background} positions social debt within the research on community smells and non-technical debt. Section~\ref{sec:methodology} describes how the six dimensions were translated into vignettes and how the ratings were analyzed. Section~\ref{sec:results} reports the rankings and tests their robustness, before Section~\ref{sec:discussion} interprets them and derives implications for practice. Section~\ref{sec:limitations} discusses limitations, and Section~\ref{sec:conclusion} concludes.

\section{Theoretical Background}
\label{sec:background}

\subsection{Social Debt and Community Smells}

Tamburri et al.\ \cite{tamburri2013} introduced social debt as the social counterpart of the technical debt metaphor, framing it as unforeseen project costs connected to a suboptimal development community. A subsequent industry study showed that engineers and managers recognize the construct even before they know the term \cite{tamburri2015}. Espinosa et al.\ \cite{espinosa2015} argued that social debt requires analysis at the individual, team, and organizational levels, and Tamburri \cite{tamburri2019} extended the construct to software architecture, locating its heaviest accrual at the boundaries where architecture and organizational structure meet.

Just as technical debt becomes visible through code smells, social debt becomes visible through \emph{community smells}: recurring patterns of dysfunctional interaction within a development community \cite{tamburri2015}. Five smells from the taxonomy consolidated by Caballero-Espinosa et al.\ \cite{caballero2023} are central to this study. An \emph{Organizational Silo} emerges when sub-teams stop sharing information, whereas a \emph{Black Cloud} describes the opposite failure, an overload of broadcast communication in which relevant information is lost. A \emph{Lone Wolf} is a contributor who works apart from the rest of the team, and a \emph{Bottleneck} is an individual who holds critical knowledge that others cannot access. \emph{Radio Silence}, finally, denotes communication breakdown in distributed or cross-cultural settings.
Caballero-Espinosa et al.\ cataloged 30 community smells across 25 studies and mapped management strategies to each. They also concluded that the field is strong on conceptualization but weak on empirical evaluation of mitigation efficacy \cite{caballero2023}, which is the observation that motivates the present study.

Zähl et al.\ \cite{zahl2026} recently organized the construct in a preliminary input-process-output model based on 24 interviews with IT professionals in the German ICT industry. \emph{Input factors} comprise antecedent conditions such as team composition, organizational structure, and individual characteristics; \emph{process dimensions} describe the mechanisms through which these antecedents turn into social debt; and \emph{output problem dimensions} describe the observable problems that mitigation measures must target. The six dimensions studied here are situated at the output level. Where a dimension shares its name with a process mechanism, as for communication breakdown and psychological unsafety, the vignettes depict the observable problem rather than the underlying mechanism.

\subsection{Antecedents and Mitigation Levers}

Several studies identify drivers of social debt. Lambiase et al.\ \cite{lambiase2022} showed in a regression model that cultural and geographical dispersion drive the emergence of Organizational Silo, Black Cloud, Lone Wolf and Radio Silence in distributed teams. Ahmad \cite{ahmad2023} identifies the absence of a psychologically safe climate as a primary upstream driver of non-technical debt in large-scale agile settings. Using machine learning on team-trait and situational variables, Zähl et al.\ \cite{zahl2026ml} found individual stress to be the strongest predictor of social debt and revealed non-linear interactions; for example, high-quality project planning offsets much of the negative effect of perceived skill imbalances within a team. Social debt is, moreover, entangled with other debt forms: Ahmad and Gustavsson \cite{ahmad2022} describe social, process, and people debt as mutually reinforcing forms of non-technical debt that also feed technical debt, and Saeeda et al.\ \cite{saeeda2024a,saeeda2024b} report that mitigation aimed at one form tends to spill over to adjacent forms and attribute most challenges in large-scale agile development to social debt.

Some of these studies also point to levers for mitigation. Catolino et al.\ \cite{catolino2019} found that gender diversity is associated with fewer community smells in open-source projects and suggest treating diversity as an active lever. Ahmad \cite{ahmad2023} positions psychological safety as both an antecedent and a target of intervention, and Dreesen et al.\ \cite{dreesen2021} propose preventive practices for distributed teams, including explicit communication norms, lower upfront interaction costs and improved psychological safety. What these contributions share is that each addresses individual dimensions or settings in isolation.

\subsection{Research Gap}

The field thus has a mature vocabulary \cite{tamburri2013,tamburri2015,espinosa2015}, a structured taxonomy of community smells \cite{tamburri2019,caballero2023}, identified antecedents \cite{lambiase2022,ahmad2023,zahl2026ml}, and an operational framework for practice \cite{zahl2026}. What is missing is a systematic, quantitative comparison of which mitigation measures practitioners perceive as most effective \emph{across} the set of social debt problem dimensions, evaluated with a common instrument \cite{caballero2023}. The present study addresses this gap.

\section{Methodology}
\label{sec:methodology}

\subsection{Research Design and Vignettes}

The study is a quantitative, cross-sectional survey following the experimental vignette methodology (EVM), specifically the paper-people variant described by Aguinis and Bradley \cite{aguinis2014}. EVM combines realism with experimental control, reduces social-desirability bias on sensitive interpersonal topics, and allows within-subject comparisons, which yield usable precision even at modest sample sizes \cite{aguinis2014}.

Six vignettes were constructed, one per problem dimension (Table~\ref{tab:dimensions}). The dimensions were derived from the output problem dimensions of Zähl et al.\ \cite{zahl2026} and matched to the community-smell taxonomy \cite{caballero2023} where a corresponding smell exists; the Psychological Safety Deficit has no counterpart in the smell taxonomy and is grounded in the leadership and non-technical-debt literature \cite{ahmad2023,dreesen2021}. Together, the six dimensions span the structural, interpersonal and cultural facets of social debt. All vignettes follow the same four-part template, which keeps the narrative form constant and varies only the target construct: a \emph{context} (a realistic German ICT team setting with team size, domain and organizational arrangement), \emph{input factors} (the trigger behind the problem, e.g., a reorganization or an acquisition), a single dominant \emph{manifestation}, and its observable \emph{consequences} (e.g., duplicated work or a failed release).

\begin{table*}[t]
\centering
\caption{Social debt problem dimensions operationalized as vignettes, corresponding community smells, number of rated measures and primary literature anchors.}
\label{tab:dimensions}
\renewcommand{\arraystretch}{1.25}
\begin{tabularx}{\textwidth}{C{0.6cm} L{3.6cm} C{1.1cm} >{\RaggedRight\arraybackslash}X L{4.0cm}}
\toprule
\textbf{ID} & \textbf{Problem dimension} & \textbf{Meas.} & \textbf{Key manifestations} & \textbf{Primary literature anchor} \\
\midrule
\rowcolor{rowgray}
V1 & Organizational Silo & 6 & Disconnected sub-teams; cross-team dependencies discovered too late & \cite{tamburri2015,caballero2023} \\
V2 & Newcomer Isolation \textit{(Lone Wolf)} & 7 & Unintegrated team member; purely transactional collaboration & \cite{tamburri2015,caballero2023} \\
\rowcolor{rowgray}
V3 & Communication Breakdown \textit{(Black Cloud)} & 7 & Channel overload; information loss; signal drowned in noise & \cite{tamburri2019,caballero2023} \\
V4 & Knowledge Bottleneck \textit{(Bottleneck)} & 7 & Knowledge concentrated in one person; stalled onboarding; bus-factor risk & \cite{tamburri2015,caballero2023} \\
\rowcolor{rowgray}
V5 & Cultural and Geographical Distance \textit{(Radio Silence)} & 7 & Distributed teams; coordination friction; mutual mistrust & \cite{lambiase2022,dreesen2021} \\
V6 & Psychological Safety Deficit & 7 & Conflict avoidance; withheld dissent; polite consensus that hides concerns & \cite{ahmad2023,dreesen2021} \\
\bottomrule
\addlinespace[2pt]
\multicolumn{5}{@{}p{\textwidth}@{}}{\footnotesize Italicized terms give the corresponding community smell where one exists. The 41 measures are listed with their results in Table~\ref{tab:results}.}
\end{tabularx}
\end{table*}

Each vignette was followed by a manipulation check asking respondents to classify the dominant problem from a uniform list of seven options (the six dimensions plus ``other/unclear''), as recommended by Aguinis and Bradley \cite{aguinis2014}. Respondents then rated six (V1) or seven (V2--V6) candidate mitigation measures, and an optional free-text item invited further measures.

\subsection{Mitigation Measures}

The 41 candidate measures were drawn from the mitigation strategies cataloged in the community-smell and social debt literature \cite{caballero2023,tamburri2015,dreesen2021,ahmad2023,saeeda2024a} and from those reported by Zähl et al.\ \cite{zahl2026}. They were phrased as concrete, actionable interventions rather than abstract goals. To avoid favoring a single type of intervention, the batteries mix three classes of measures: \emph{structural or process measures} (S), which change how work, information, or decisions are organized (e.g., single-channel consolidation, decision logging); \emph{role or ownership measures} (R), which change who is responsible for what (e.g., rotation, mentoring, dedicated owners); and \emph{attitudinal or training measures} (A), which aim to change awareness, skills, or attitudes (e.g., communication-skills training, psychological-safety workshops). Each measure was assigned to exactly one class (Table~\ref{tab:results}). Five of the six batteries contain measures from all three classes; the V4 battery contains no attitudinal measures.

\subsection{Sample and Procedure}

The target population comprised IT professionals and advanced IT students or recent graduates ($\leq$2~years since graduation). Respondents were recruited through professional and academic networks using convenience and snowball sampling; participation was voluntary, anonymous, and not incentivized. Group membership (professional vs.\ student) was recorded to allow the two subsamples to be compared. Including students is defensible for a judgment task of this kind: H\"{o}st et al.\ \cite{host2000} found only minor differences between students and professionals in a software engineering judgment task, Salman et al.\ \cite{salman2015} found comparable performance when both groups applied an approach that was new to them, and expert commentaries caution against dismissing student samples in general \cite{falessi2018,feldt2018}. Social debt was a new concept for most respondents, which puts both groups on a similar footing.

The questionnaire was implemented in Google Forms and comprised demographic and control items, the six vignettes, and two closing items on the perceived seriousness of social debt and the maturity of formal practices in the respondent's organization, plus optional free-text reflections. It was open for about two weeks and took roughly ten minutes to complete. Respondents were informed about the purpose of the study, anonymity, and their right to withdraw. The instrument was reviewed within the research team, which led to minor wording changes. Scenarios and measures were presented in a fixed order. Only complete questionnaires were analyzed. Perceived efficacy was rated on a five-point scale (1~=~very ineffective, 5~=~very effective). Parametric analysis of such ratings is robust in practice \cite{norman2010}; an ordinal model was fitted as a robustness check (Section~\ref{sec:analysis}).

\subsection{Analysis}
\label{sec:analysis}

The analysis focuses on within-scenario rankings because the research question asks which measures are preferred over their alternatives. All analyses were run in Python; data and code are available in the appendix.

\textit{Descriptive and frequentist analysis.} For each measure, we computed mean, standard deviation, and median, and ranked the measures within each scenario. One-sample $t$-tests compared each scenario's top-ranked measure with the neutral midpoint, and paired $t$-tests compared it with every alternative in the same scenario (effect size $d_z$). Because the top measure is selected on the same data and 35 such contrasts were run, we report Holm-adjusted $p$-values \cite{holm1979} alongside unadjusted ones. Manipulation checks were tabulated per scenario, and a sensitivity analysis re-computed all rankings without respondents who failed the respective check.

\textit{Bayesian hierarchical model.} Ranking by raw means treats each mean as known without error and overstates confidence in whichever measure happens to rank first. We therefore fitted a hierarchical model in which the rating of respondent $i$ for measure $k$ in scenario $v$ is
\begin{equation}
y_{ik} = \mu_v + \beta_k + u_{iv} + \varepsilon_{ik}, \quad \beta_k \sim \mathcal{N}(0,\tau_v^2), \quad u_{iv} \sim \mathcal{N}(0,\sigma_{u,v}^2).
\label{eq:model}
\end{equation}
The measure effects $\beta_k$ are partially pooled toward the scenario mean $\mu_v$, so estimates are shrunk in proportion to how little the data distinguish the measures; $\tau_v$ quantifies how much the measures in a scenario differ. The respondent intercepts $u_{iv}$ preserve the within-subject structure, which a model of the measure means alone would ignore. Priors were weakly informative ($\mu_v \sim \mathcal{N}(3.5, 1)$, $\tau_v \sim \text{Half-}\mathcal{N}(0.5)$, $\sigma_{u,v}, \sigma_v \sim \text{Half-}\mathcal{N}(1)$). The model was fitted in PyMC \cite{pymc2023} with four chains of 2,000 draws after 2,000 tuning steps (all $\hat{R} \leq 1.003$, no divergent transitions). For each measure we report the posterior mean, a 95\% credible interval (CrI) and $P(\text{best})$, the posterior probability that it has the highest efficacy in its scenario. With six or seven measures, a $P(\text{best})$ of about $0.14-0.17$ corresponds to no preference. As a robustness check, the same model was fitted with a cumulative-logit (ordinal) likelihood.

\textit{Class-level test.} To test whether the preference for structural measures holds across scenarios rather than only for individual winners, we extended the model with fixed effects for the three measure classes and a random effect for each measure within its class, so that the class contrast accounts for the heterogeneity of individual items; respondent-specific slopes allow individuals to differ in their preference for structural over attitudinal measures. As a simpler complement, we compared per-respondent class means with paired $t$-tests.

\textit{Further analyses.} Professionals and students were compared on overall and scenario-level mean efficacy (Welch $t$-test, Mann-Whitney $U$), and Spearman correlations related to experience, familiarity with social debt, and the closing items. Given subsamples of $n=8$ and $n=13$, these comparisons are exploratory.

\section{Results}
\label{sec:results}

\subsection{Sample and Manipulation Checks}

Twenty-three respondents completed the survey. Of these, 13 (56.5\%) were students, 8 (34.8\%) were professionals, and 2 (8.7\%) did not state their role. The sample was predominantly male (19; 82.6\%) and early-career: 5 respondents (21.7\%) had less than one year of IT experience, 15 (65.2\%) one to three years and 3 (13.0\%) four to seven years. Most worked in small teams (1-5 people: 43.5\%; 6--10: 30.4\%; 11--20: 21.7\%; 21+: 4.3\%) and in an office-based hybrid arrangement (52.2\%; fully remote 26.1\%, remote-leaning hybrid 13.0\%, co-located 8.7\%). Only 4 respondents (17.4\%) were familiar with the term social debt; 12 (52.2\%) had heard of it and 7 (30.4\%) encountered it for the first time.

Respondents identified the intended dimension in 95.7\% of cases for V1, V4 and V6, 91.3\% for V3 and 82.6\% for V5. Newcomer Isolation (V2) was recognized least often (69.6\%); of the seven misclassifications, four named the Psychological Safety Deficit and two the Organizational Silo, indicating conceptual overlap with these dimensions. Excluding respondents who failed the respective check left the rankings in V1 and V3-V6 unchanged. In V2 ($n=16$), the two leading measures swapped places (team norms around code review and pairing: $M=3.94$; cross-functional projects: $M=3.88$) and no measure moved by more than one rank.

\subsection{Rankings of Mitigation Measures}

\begin{figure*}[htb]\centering
\includegraphics[width=\textwidth,trim={0.2cm 0.0cm 0cm 0.0cm},clip]{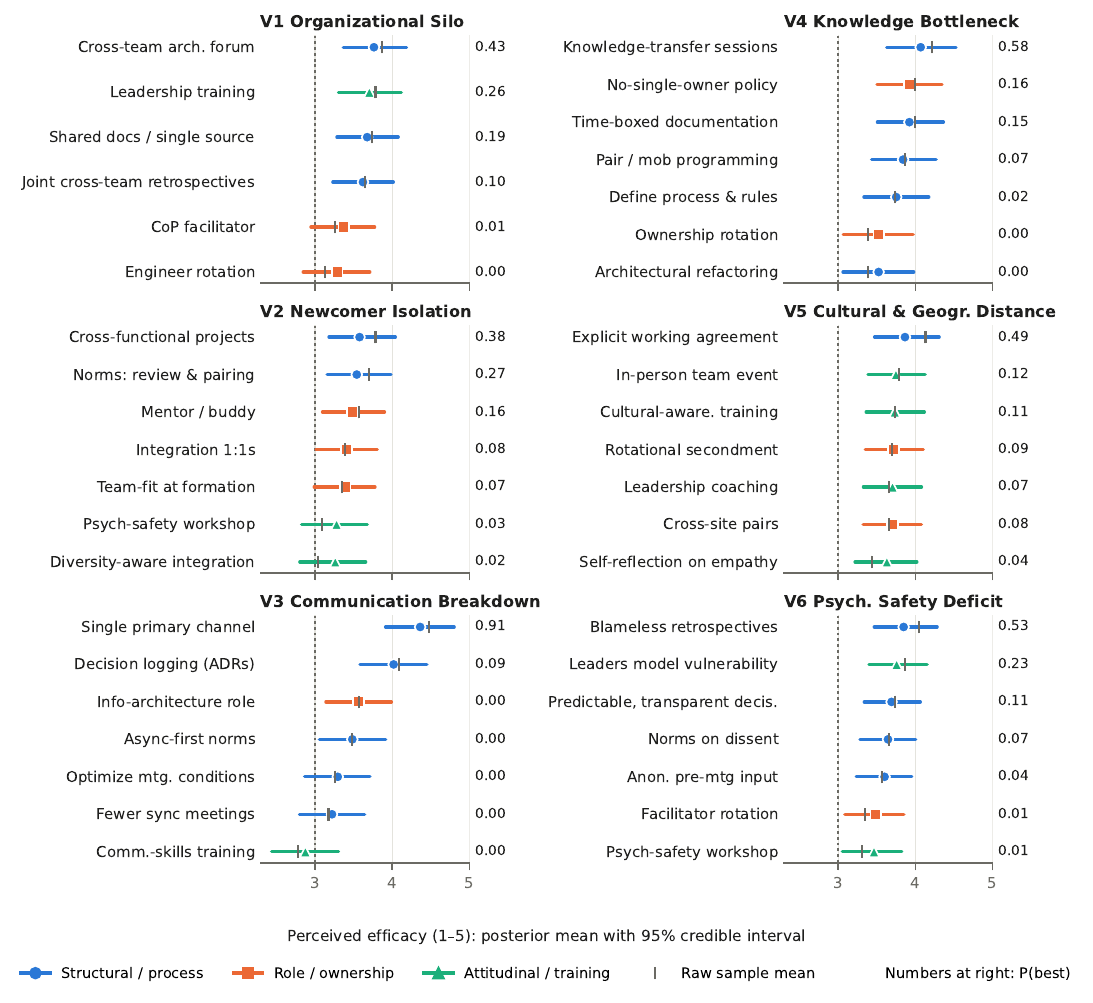}
\caption{Posterior mean perceived efficacy and 95\% credible intervals for all 41 measures, colored and shaped by measure class; vertical ticks show raw means, the dotted line marks the scale midpoint. Numbers at the right give the posterior probability that a measure is the most effective in its scenario.}
\label{fig:forest}
\end{figure*}

Table~\ref{tab:results} reports all 41 measures, and Fig.~\ref{fig:forest} shows the posterior estimates. In every scenario, the top-ranked measure was rated clearly above the neutral midpoint (one-sample $t(22)$ between 3.23 and 6.83, all $p \leq 0.004$, $d$ between 0.67 and 1.42), so the leading measures are endorsed in absolute terms and not merely relative to weaker alternatives.

\begin{table*}[!htb]\centering\scriptsize
\caption{Perceived efficacy of all 41 mitigation measures ($N=23$), ordered by raw mean within scenario. Cl.\ = measure class (S structural/process, R role/ownership, A attitudinal/training). Post.\ $M$ and 95\% CrI from the hierarchical model (Eq.~\ref{eq:model}); $\tau_v$ = posterior mean [95\% CrI] of the between-measure SD. $d_z$ and $p$ (Holm-adjusted $p$ in parentheses) refer to the paired contrast with the scenario's top measure; Holm adjustment across all 35 contrasts.}
\label{tab:results}
\begin{tabularx}{\textwidth}{@{}>{\RaggedRight\arraybackslash}X c c c c c c@{}}
\toprule
\hdr{Mitigation measure} & \hdr{Cl.} & \hdr{$M$ ($SD$)} & \hdr{Post.\ $M$ [95\% CrI]} & \hdr{$P$(best)} & \hdr{$d_z$} & \hdr{$p$ ($p_\text{Holm}$)} \\
\midrule
\multicolumn{7}{@{}l}{\textbf{V1 Organizational Silo} \quad { $\tau_v=0.29$ [0.05, 0.64]}} \\
\textbf{Establish a regular cross-team architecture forum} & S & 3.87 (1.29) & 3.76 [3.37, 4.19] & \textbf{0.43} & -- & -- \\
\rowcolor{rowgray}Provide leadership training on cross-team coordination & A & 3.78 (1.04) & 3.70 [3.30, 4.12] & 0.26 & 0.06 & 0.792 (1.000) \\
Implement shared technical documentation standards and a single source of truth & S & 3.74 (1.05) & 3.68 [3.28, 4.08] & 0.19 & 0.11 & 0.589 (1.000) \\
\rowcolor{rowgray}Conduct joint cross-team retrospectives & S & 3.65 (1.11) & 3.62 [3.23, 4.01] & 0.10 & 0.16 & 0.447 (1.000) \\
Appoint a dedicated community-of-practice facilitator & R & 3.26 (0.75) & 3.37 [2.95, 3.77] & 0.01 & 0.57 & 0.013 (0.331) \\
\rowcolor{rowgray}Introduce rotation of engineers across teams (e.g., quarterly) & R & 3.13 (0.92) & 3.29 [2.85, 3.71] & 0.00 & 0.57 & 0.012 (0.312) \\
\midrule
\multicolumn{7}{@{}l}{\textbf{V2 Newcomer Isolation} \quad { $\tau_v=0.21$ [0.01, 0.55]}} \\
\textbf{Include in cross-functional, collaborative projects requiring coordination} & S & 3.78 (0.95) & 3.58 [3.18, 4.04] & \textbf{0.38} & -- & -- \\
\rowcolor{rowgray}Explicit team norms around code review and pair-programming participation & S & 3.70 (1.06) & 3.54 [3.16, 3.98] & 0.27 & 0.06 & 0.765 (1.000) \\
Assign a mentor or buddy from outside their immediate technical domain & R & 3.57 (1.34) & 3.48 [3.10, 3.90] & 0.16 & 0.24 & 0.260 (1.000) \\
\rowcolor{rowgray}Structured 1:1s with the engineering manager focused on team integration & R & 3.39 (1.41) & 3.41 [3.01, 3.81] & 0.08 & 0.27 & 0.215 (1.000) \\
Ensure compatibility when creating the team & R & 3.35 (1.11) & 3.39 [2.99, 3.78] & 0.07 & 0.32 & 0.135 (1.000) \\
\rowcolor{rowgray}Psychological-safety-building team workshop & A & 3.09 (1.16) & 3.28 [2.83, 3.68] & 0.03 & 0.55 & 0.015 (0.351) \\
Diversity-aware team integration practices & A & 3.04 (1.19) & 3.26 [2.80, 3.66] & 0.02 & 0.56 & 0.014 (0.341) \\
\midrule
\multicolumn{7}{@{}l}{\textbf{V3 Communication Breakdown} \quad { $\tau_v=0.56$ [0.31, 0.95]}} \\
\textbf{Consolidate communication into a single primary channel with clear conventions} & S & 4.48 (1.04) & 4.36 [3.91, 4.80] & \textbf{0.91} & -- & -- \\
\rowcolor{rowgray}Establish explicit decision-logging practices (e.g., Architecture Decision Records) & S & 4.09 (1.00) & 4.02 [3.58, 4.45] & 0.09 & 0.42 & 0.059 (1.000) \\
Create a dedicated information-architecture role (team scribe / documentation owner) & R & 3.57 (1.16) & 3.56 [3.14, 3.99] & 0.00 & 0.66 & 0.004 (0.127) \\
\rowcolor{rowgray}Adopt asynchronous-first communication norms with response-time expectations & S & 3.48 (0.90) & 3.48 [3.06, 3.91] & 0.00 & 1.00 & $<$0.001 (0.003) \\
Optimize framework conditions (working conditions, number of meetings) & S & 3.26 (1.05) & 3.29 [2.87, 3.71] & 0.00 & 0.90 & $<$0.001 (0.008) \\
\rowcolor{rowgray}Reduce synchronous meetings, replacing them with structured async updates & S & 3.17 (1.23) & 3.22 [2.80, 3.64] & 0.00 & 1.18 & $<$0.001 ($<$0.001) \\
Provide communication-skills training for the team & A & 2.78 (0.95) & 2.88 [2.44, 3.30] & 0.00 & 1.35 & $<$0.001 ($<$0.001) \\
\midrule
\multicolumn{7}{@{}l}{\textbf{V4 Knowledge Bottleneck} \quad { $\tau_v=0.30$ [0.07, 0.62]}} \\
\textbf{Structured knowledge-transfer sessions with the bottleneck engineer} & S & 4.22 (1.13) & 4.07 [3.63, 4.53] & \textbf{0.58} & -- & -- \\
\rowcolor{rowgray}Explicit `no-single-owner' policy enforced via code-ownership rules & R & 4.00 (1.09) & 3.92 [3.50, 4.34] & 0.16 & 0.18 & 0.410 (1.000) \\
Time-boxed documentation initiative with management commitment & S & 4.00 (0.90) & 3.92 [3.51, 4.36] & 0.15 & 0.22 & 0.308 (1.000) \\
\rowcolor{rowgray}Pair-programming or mob-programming on the affected subsystems & S & 3.87 (1.01) & 3.84 [3.43, 4.27] & 0.07 & 0.37 & 0.088 (1.000) \\
Define process \& rules (communication points, collaboration, process model / agility) & S & 3.74 (1.29) & 3.75 [3.33, 4.17] & 0.02 & 0.36 & 0.102 (1.000) \\
\rowcolor{rowgray}Architectural refactoring to reduce subsystem complexity & S & 3.39 (0.84) & 3.52 [3.06, 3.98] & 0.00 & 1.06 & $<$0.001 (0.001) \\
Scheduled rotation of subsystem ownership & R & 3.39 (1.34) & 3.53 [3.07, 3.97] & 0.00 & 0.65 & 0.005 (0.137) \\
\midrule
\multicolumn{7}{@{}l}{\textbf{V5 Cultural and Geographical Distance} \quad { $\tau_v=0.15$ [0.01, 0.42]}} \\
\textbf{Explicit shared working agreement (hours, response times, decision norms)} & S & 4.13 (1.10) & 3.86 [3.47, 4.31] & \textbf{0.49} & -- & -- \\
\rowcolor{rowgray}In-person team-building event with all sides present & A & 3.78 (1.35) & 3.75 [3.39, 4.13] & 0.12 & 0.37 & 0.088 (1.000) \\
Cultural-awareness training for all sides & A & 3.74 (0.81) & 3.73 [3.36, 4.12] & 0.11 & 0.31 & 0.154 (1.000) \\
\rowcolor{rowgray}Rotational secondment (engineers working temporarily at the other site) & R & 3.70 (1.15) & 3.72 [3.35, 4.10] & 0.09 & 0.33 & 0.125 (1.000) \\
Establish cross-site engineer pairs for ongoing 1:1 collaboration & R & 3.65 (1.15) & 3.70 [3.32, 4.08] & 0.08 & 0.48 & 0.031 (0.678) \\
\rowcolor{rowgray}Leadership coaching for site leads on cross-cultural collaboration & A & 3.65 (1.15) & 3.70 [3.33, 4.08] & 0.07 & 0.38 & 0.086 (1.000) \\
Assess your own reticence/empathy and expectations of other team members & A & 3.43 (1.04) & 3.63 [3.22, 4.02] & 0.04 & 0.54 & 0.017 (0.395) \\
\midrule
\multicolumn{7}{@{}l}{\textbf{V6 Psychological Safety Deficit} \quad { $\tau_v=0.22$ [0.02, 0.53]}} \\
\textbf{Structured retrospectives using `blameless' facilitation protocols} & S & 4.04 (1.07) & 3.85 [3.47, 4.28] & \textbf{0.53} & -- & -- \\
\rowcolor{rowgray}Leaders explicitly modeling vulnerability & A & 3.87 (0.97) & 3.76 [3.40, 4.15] & 0.23 & 0.18 & 0.406 (1.000) \\
Enable predictability and create transparency in decision-making processes & S & 3.74 (0.92) & 3.69 [3.34, 4.06] & 0.11 & 0.28 & 0.200 (1.000) \\
\rowcolor{rowgray}Explicit team norms around dissent (e.g., `disagree and commit') & S & 3.65 (1.03) & 3.64 [3.28, 4.00] & 0.07 & 0.35 & 0.107 (1.000) \\
Anonymous pre-meeting decision-input mechanism (e.g., written-first discussions) & S & 3.57 (1.04) & 3.60 [3.23, 3.95] & 0.04 & 0.35 & 0.110 (1.000) \\
\rowcolor{rowgray}Rotation of facilitator role within team meetings & R & 3.35 (0.88) & 3.49 [3.09, 3.85] & 0.01 & 0.47 & 0.036 (0.748) \\
Explicit psychological-safety workshop & A & 3.30 (1.18) & 3.46 [3.05, 3.82] & 0.01 & 0.81 & $<$0.001 (0.025) \\
\bottomrule
\end{tabularx}
\end{table*}

The rankings differ considerably in how decisive they are. Communication Breakdown (V3) has the only clear winner: consolidating communication into a single primary channel reaches $P(\text{best})=0.91$, the measures in this scenario differ the most ($\tau_v=0.56$), and the top measure outperforms four of its six alternatives even after Holm adjustment, with communication-skills training rated the lowest of all 41 measures ($M=2.78$; $d_z=1.35$). Only decision logging comes close ($M=4.09$, $p=0.059$).

In the other five scenarios, the raw-mean winner remains the most likely best option under the hierarchical model, but with moderate certainty: structured knowledge transfer in V4 ($P(\text{best})=0.58$), blameless retrospectives in V6 (0.53), an explicit shared working agreement in V5 ($0.49$), a cross-team architecture forum in V1 ($0.43$) and inclusion in cross-functional projects in V2 (0.38). Each of these probabilities is roughly two and a half to four times the chance level of $0.14-0.17$, but none rules out the runner-up. Accordingly, few paired contrasts survive the Holm adjustment outside V3: knowledge transfer versus architectural refactoring in V4 ($p_\text{Holm}=0.001$) and blameless retrospectives versus a stand-alone psychological-safety workshop in V6 ($p_\text{Holm}=0.025$). In V1, the top four measures are practically indistinguishable (paired $d_z \leq 0.16$) and separate only from the two role-based measures, rotation and a community-of-practice facilitator. V2 is the least decisive scenario: its two leading measures differ by less than 0.1 scale points and swap places in the sensitivity analysis. V5 shows the smallest between-measure variation ($\tau_v=0.15$, 95\% CrI [0.01, 0.42]); the working agreement leads, but six alternatives lie within 0.2 posterior points of each other.

The ordinal model reproduced these conclusions: the same measure ranked first in all six scenarios, with $P(\text{best})$ between 0.42 (V2) and 0.96 (V3).

\subsection{Structural Versus Attitudinal Measures}

Across scenarios, structural measures were rated higher than both other classes. In the hierarchical class model, structural measures exceeded attitudinal measures by 0.39 scale points (95\% CrI [0.09, 0.68]; $P(>0)=0.99$) and role-based measures by 0.32 points ([0.06, 0.56]; $P(>0)=0.99$), whereas role-based and attitudinal measures did not differ (0.07, [$-$0.22, 0.36]). The per-respondent class means lead to the same conclusion (structural $M=3.78$, role $3.49$, attitudinal $3.45$; S vs.\ A: $t(22)=4.74$, $p<0.001$, $d_z=0.99$; S vs.\ R: $t(22)=3.87$, $p<0.001$, $d_z=0.81$).

The pattern is not uniform across scenarios. The structural-attitudinal gap was pronounced in V3 (0.91 points, $p<0.001$), V2 (0.67, $p=0.002$) and V5 (0.48, $p=0.014$), small in V6 (0.16, $p=0.24$) and absent in V1 ($-$0.03, $p=0.90$), where leadership training on cross-team coordination ranked second. V4 contains no attitudinal measure; there, structural knowledge transfer was leading, and the two role-based measures ranked second (a no-single-owner policy) and last (ownership rotation).

\subsection{Group Comparison, Closing Items and Free Text}

Professionals rated the measures somewhat higher overall than students ($M=3.96$ vs.\ $3.54$; Welch $t=1.87$, $p=0.077$, $d=0.77$; Mann-Whitney $U=68.0$, $p=0.26$). At scenario level, the difference was significant for V3 ($d=1.11$, $p=0.013$) and V4 ($d=0.96$, $p=0.023$), but neither survives a Holm adjustment across the nine comparisons. The groups agreed moderately on the \emph{order} of the 41 measures (Spearman $\rho=0.52$, $p<0.001$). Agreement was high for V3 ($\rho=0.91$) and moderate for V1, V4 and V6 ($\rho$ between 0.57 and 0.68), but low for V2 ($\rho=0.20$) and absent for V5 ($\rho=-0.06$), where professionals rated the in-person team event highest. Both groups rated structural measures above attitudinal measures (professionals 4.11 vs.\ 3.72; students 3.70 vs.\ 3.39).

Respondents considered social debt a moderately serious problem ($M=3.55$, $SD=0.96$) and rated the formal practices in their organizations slightly lower ($M=3.36$, $SD=1.14$; $n=22$ for both items, as one respondent skipped the closing section). Spearman correlations between seriousness and formal practices ($\rho=0.15$, $p=0.51$), experience and overall efficacy ($\rho=-0.10$, $p=0.64$) and familiarity and seriousness ($\rho=-0.22$, $p=0.33$) were weak and non-significant, as were the group differences on the closing items (both $p \geq 0.40$).

The optional free-text answers were not systematically coded and serve only as illustrations. They mostly echoed the top-rated measures, calling for clear accountability (dependency owners, succession and backup plans), explicit communication conventions with a single source of truth, structured conversations for integrating newcomers, and leaders who actively invite dissent.

\section{Discussion}
\label{sec:discussion}

\subsection{Interpretation of Findings}

Three findings answer the research question. First, in each scenario, one measure leads the ranking, but the confidence in these leaders varies widely: only for Communication Breakdown is the answer unambiguous, while for the other five dimensions, the data identify a favorite with moderate evidence and a small group of close alternatives. Second, across scenarios, respondents prefer measures that change the \emph{structure} of how a team works, that is, where information lives, how decisions are recorded, and how knowledge flows, over measures that assign roles or aim to change attitudes. Third, role-based measures such as rotation, dedicated facilitators, or ownership rules were rated no better than training-based ones; the preference is specifically for structural change, not simply for ``hard'' over ``soft'' measures.

This pattern is consistent with the socio-technical framing of social debt, in which communication infrastructure and organizational topology are primary leverage points \cite{tamburri2019}, and with Dreesen et al.\ \cite{dreesen2021}, whose most actionable recommendations concern explicit communication norms and lower interaction costs. The V3 result illustrates this most clearly. The scenario describes a communication problem, yet, communication-skills training was the lowest-rated measure in the entire study. Respondents apparently separated the symptom from its cause: the vignette describes too many channels and lost information, a problem of how communication is organized rather than how well individuals communicate. Measures that address this organization directly, consolidating channels and logging decisions, were rated highest. Similarly, a stand-alone psychological-safety workshop ranked last in V6 and was clearly outperformed by blameless retrospectives. This does not suggest that psychological safety is unimportant; rather, respondents seem to regard it as an outcome of structures and leadership behavior, in line with Ahmad \cite{ahmad2023}, who places psychological safety downstream of leadership.

The exceptions qualify this picture. In the Organizational Silo scenario (V1), leadership training ranked second, and the top four measures, which include forums, shared documentation, joint retrospectives, and training, were practically indistinguishable. Silos appear to be seen as a problem with several complementary entry points rather than a single fix. In V6, leaders explicitly modeling vulnerability ranked second, which suggests that leadership \emph{behavior} is valued even where leadership \emph{training} in general is not. Mitigation advice for these dimensions is therefore better framed as a portfolio than as a single recommended action.

\subsection{Relation to Prior Work and Alternative Explanations}

The study responds to the gap identified by Caballero-Espinosa et al.\ \cite{caballero2023} and translates the qualitative dimensions of Zähl et al.\ \cite{zahl2026} into a quantitative, comparative prioritization. The high manipulation-check rates for five of six scenarios indicate that the vignettes isolated their intended dimensions; the weaker result for Newcomer Isolation and its confusions with the Psychological Safety Deficit and the Organizational Silo suggest that newcomer integration is perceived as a blend of interpersonal and structural problems.

Several alternative explanations for the preference for structural measures deserve consideration. \emph{Social desirability}, i.e., the perception that structural measures merely sound more professional, is weakened by the fact that some structural measures were rated low (e.g., architectural refactoring in V4, fewer synchronous meetings in V3); respondents discriminated within the class. A \emph{menu artifact} is unlikely, since five of six batteries contained all three classes, although the specific measures chosen per class may still have influenced the outcome. \emph{Acquiescence or central tendency} is contradicted by the spread of means within scenarios (up to 1.7 scale points in V3). A \emph{wording effect} is harder to exclude: structural measures may have been phrased more concretely (``consolidate communication into a single primary channel'') than attitudinal ones (``communication-skills training''), and concreteness alone can raise perceived actionability. All measures were phrased as concrete interventions, but future work should explicitly control item wording and length.

\subsection{Implications for Practice}

For engineering managers, the results offer a first-response heuristic, graded by the strength of the evidence. When communication overloads a team (V3), consolidating channels and logging decisions are the clear first choices, well ahead of communication-skills training. When a single engineer becomes a bottleneck (V4), structured knowledge transfer, backed by a time-boxed documentation effort, is the most promising starting point, ahead of rotating ownership. When teams stay silent (V6), blameless retrospectives and visible leadership modeling are preferable to stand-alone workshops. For distributed teams (V5), an explicit shared working agreement is the most likely favorite, but in-person events, cultural-awareness training, and cross-site pairing are close alternatives, and professionals favored in-person contact. For silos (V1) and newcomer integration (V2), several measures are about equally promising; a cross-team architecture forum and real cross-functional project work, respectively, are reasonable starting points, combined with complementary measures from the leading group. The overarching lesson is that defining where information lives, how decisions are recorded, and how knowledge is transferred tends to be perceived as higher-leverage than rotating roles or training attitudes. As these are perceptions rather than measured outcomes, the heuristic narrows the space of plausible interventions but does not replace local judgment.

\section{Limitations}
\label{sec:limitations}

\textit{Statistical conclusion validity.} The realized sample ($N=23$) fell short of the planned 30-50. Because every respondent rated every measure, the within-scenario contrasts are paired; at $N=23$ they detect effects of $d_z \geq 0.61$ with 80\% power ($\alpha=0.05$, two-sided). Medium-sized differences between measures, such as most contrasts outside V3, therefore remain uncertain, which the hierarchical model reflects in moderate $P(\text{best})$ values. The group comparison could only detect very large effects ($d \geq 1.33$ with $n=8$ and $n=13$) and cannot establish that professionals and students judge the measures alike.

\textit{Construct validity.} The dependent variable is \emph{perceived} efficacy. Perceptions matter because they shape which measures are adopted, but they are no substitute for outcome data such as delivery performance or collaboration quality. Newcomer Isolation had the weakest manipulation check, and its ranking was the least stable. The assignment of measures to classes involves judgment, and some measures (e.g., leaders modeling vulnerability) could be classified differently.

\textit{External validity.} The sample is small, mostly male (82.6\%), early-career (86.9\% with at most three years of experience) and student-heavy (56.5\%), and was recruited through convenience and snowball sampling. The findings, therefore, describe how IT practitioners and trainees perceive these measures rather than how seasoned professionals do; the modest rank agreement between the subgroups, especially for V2 and V5, underlines the need for a professionals-only replication. The gender imbalance is a particular concern for a study of social dynamics, given evidence that team diversity affects community health \cite{catolino2019}. The German ICT setting of the vignettes further constrains transfer to other contexts.

\textit{Design and content.} Scenarios and measures were presented in a fixed order, so order and fatigue effects cannot be ruled out. The measures were literature-derived and necessarily non-exhaustive; free-text suggestions such as dedicated dependency owners and succession plans should be included as fixed options in future work. Finally, each vignette isolates one dominant problem, whereas in practice, social debt manifestations co-occur and interact.

\section{Conclusion}
\label{sec:conclusion}

This study translated six social debt problem dimensions into vignettes and asked 23 IT professionals and advanced students to rate 41 literature-derived mitigation measures. A hierarchical Bayesian model that accounts for the within-subject design shows that the leading measure is unambiguous only for Communication Breakdown, where consolidating communication into a single channel is almost certainly preferred. For the other five dimensions, it identifies favorites with moderate evidence: structured knowledge transfer, blameless retrospectives, an explicit working agreement for distributed teams, a cross-team architecture forum, and cross-functional project work for newcomers. Across all scenarios, structural measures were rated 0.3 to 0.4 scale points higher than role-based or attitudinal ones, a pattern that was strong for communication, newcomer, and distance problems but absent for silos.

The results offer engineering managers an evidence-graded starting point for choosing among mitigation measures. Given the small, convenience-based sample, they should be treated as hypotheses. Confirmatory studies with larger, professionals-only samples, randomized presentation, and controlled item wording should test them and ultimately move from perceived to measured effects on team and project outcomes.


\section*{Appendix}
All additional materials (questionnaire, source files, data) are available at: \url{https://doi.org/10.7910/DVN/SAKK5P}


\end{document}